\documentstyle[aps]{revtex}
\begin{document}
\centerline{\Large {\bf Two Magnetic Impurities in a Spin Chain}}
\centerline{\bf Zhan-Ning Hu\footnote{{\bf 
E-mail: huzn@aphy.iphy.ac.cn}} and Fu-Cho Pu\footnote{{\bf E-mail:
pufc@aphy.iphy.ac.cn}}}\centerline{Institute of Physics and Center for
Condensed Matter Physics,} \centerline{Chinese Academy of Sciences, Beijing
100080, China}

\begin{center}
\begin{minipage}{5in}
\centerline{\large\bf   Abstract}
In this letter, the Kondo magnetic effect is studied for the $XXZ$ 
spin chain where the impurities are coupled to the edges of the 
system. The Hamiltonian of our model can be constructed from the transfer 
matrix. It is exactly solvable and the exchange constants between the bulk 
and the boundary impurities are arbitrary. The finite size corrections to 
the ground state energy are obtained for the magnetic impurities and the 
boundary condition. The specific heat and the susceptibility contributed 
by the impurities are derived and the Kondo temperature is given 
explicitly by the use of the Luttinger-Fermi liquid picture and 
the Bethe ansatz method.



\smallskip

\end{minipage}
\end{center}

\newpage
%%\raggedcolumns
%%\begin{multicols}{2}
%%\narrowtext

It is well known that the spin dynamics of the Kondo problem is equivalent
to the dynamics of the spin chain with the magnetic impurities\cite{yp6}.
Much progress has been made recently on the impurity models based on the
methods of renormalization-group techniques\cite{che3}, conformal field
theory\cite{che4,a001,a002} and the integrability investigations. The
quantum inverse scattering method and the Bethe ansatz technique have been
the powerful tools to study the integrable impurities within the framework
of the quantum spin chains.

Magnetic impurities play an important role in the strongly correlated
electron systems, especially for one dimensional systems. And the
impurity usually destroys the integrability when it is embedded into the
`pure' quantum chain. Then it is a challenging problem to deal with the
impurity effects in the strongly correlated electron systems for the quantum
integrable models without losing the integrability. Now several integrable
impurity models have been found\cite{many01,many02,many03,many04} besides
the magnetic impurities in the noninteracting metals such as the Anderson
impurity model and the multichannel Kondo problem\cite{yp6,2701,2702,2703}.

As a quantum mechanical model of magnetism, the Heisenberg spin chain turned
out to be very fruitful and the integrable impurity problem for this model
was first considered in Ref.[13] and generalized to arbitrary spin in
Ref. [14]. Besides the natural understanding of completely integrable
quantum systems, the quantum method of the inverse problem provides us also a
very useful technique to construct the impurity model which preserves 
integrability. Indeed, this scheme has been adapted in Ref. [6-8]. 
But, for the periodic boundary systems, some
unphysical terms must be present in the Hamiltonian to maintain the 
integrability, though they may be irrelevant when the impurities are
embedded in the system. Notice that the fact that the impurities cut the
one-dimensional system into the `piece' when they are introduced and (then)
the open boundary systems are formed with the impurities at the ends of the
systems. So the integrable impurity models\cite{many05,many06,many07,many08}
can be studied with the use of open boundary conditions. In recent years,
various integrable models on the open chains with boundary terms have been
investigated (See, for example, references [19-22]). 
In Ref. [23] the open Heisenberg chain with impurities has been
discussed and the integrability is given in Ref. [24] when the two
impurities are coupled to the $XXZ$ chain. In the present letter, we will
discuss the effects of the magnetic impurities in the $XXZ$ spin chain and
the finite size correction contributed by the impurities. Our starting point
is the following Hamiltonian

\begin{eqnarray}
H &=&\sum_{j=1}^{N-1}\sigma _j^x\sigma _{j+1}^x+\sigma _j^y\sigma
_{j+1}^y+\cosh \eta \sigma _j^z\sigma _{j+1}^z  \nonumber \\
&&+J\left( \sigma _1^xd_a^x+\sigma _1^yd_a^y+\frac{\cosh \eta }{\cosh c}
\sigma _1^zd_a^z\right)  \label{s002} \\
&&+J\left( \sigma _N^xd_b^x+\sigma _N^yd_b^y+\frac{\cosh \eta }{\cosh c}
\sigma _N^zd_b^z\right)  \nonumber
\end{eqnarray}
where the constants $J,$ $c$ and $\eta $ satisfy the relation $J=\sinh
^2\eta \cosh c/\left( \sinh ^2\eta -\sinh ^2c\right) $ . The Pauli operators 
$\sigma _j^{x,y,z}$ act non-trivially only in the $j$th quantum vector space
and $d_a^{x,y,z}$, $d_b^{x,y,z}$ are the spins of magnetic impurities with $
s=1/2$ . This system can been constructed from the transfer matrix\cite
{boook} 
\begin{equation}
U\left( \lambda \right) =Tr_\tau \left\{ K^{+}\left( \lambda \right) T_\tau
\left( \lambda \right) K^{-}\left( \lambda \right) \widehat{T}_\tau \left(
\lambda \right) \right\}  \label{s003}
\end{equation}
where 
\[
T_\tau \left( \lambda \right) =L_{\tau b}\left( \lambda +c\right) L_{\tau
N}\left( \lambda \right) \cdots L_{\tau 2}\left( \lambda \right) L_{\tau
1}\left( \lambda \right) , 
\]
\[
\widehat{T}_\tau \left( \lambda \right) =L_{\tau 1}\left( \lambda \right)
L_{\tau 2}\left( \lambda \right) \cdots L_{\tau N}\left( \lambda \right)
L_{\tau b}\left( \lambda -c\right) .
\]
The $L$ operator is given by  
\[
L_{\tau n}\left( \lambda \right) =\sum_{j=1}^4w_j\sigma _\tau ^j\otimes
\sigma _n^j 
\]
with 
\[
w_1=w_2=\frac 12\sinh \eta , 
\]
\[
w_4-w_3=\sinh \lambda , 
\]
\[
w_4+w_3=\sinh \left( \lambda +\eta \right) 
\]
for the subindex $n=1,2,\cdots ,N$ $,a,b$ being the quantum vector spaces
and the impurity sites at both of the ends. The symbol $Tr$ denotes the
trace in the auxiliary space $\tau $ in $C^2$. The $\sigma ^j$ are the Pauli
operators for $j=1,2,3$ with $\sigma _j^4$ being the identity operator. The
boundary reflection matrices take their forms as 
\[
K^{+}\left( \lambda \right) =I,\;K^{-}\left( \lambda \right) =L_{\tau
a}\left( \lambda +c\right) L_{\tau a}\left( \lambda -c\right) 
\]
where $c$ is an arbitrary constant. Then the Hamiltonian (\ref{s002}) can be
obtained from the logarithmic derivative of the transfer matrix $U\left(
\lambda \right) $ when the spectral parameter $\lambda \rightarrow 0$ . By
the use of the standard Bethe ansatz method, the 
eigenvalue of the Hamiltonian is 
\begin{eqnarray*}
E &=&4\sinh \eta \sum_{j=1}^M\frac{\sinh \eta }{\cosh \left( 2\lambda
_j\right) -\cosh \eta } \\
&&+\left( N-1+\frac{2J}{\cosh c}\right) \cosh \eta
\end{eqnarray*}
for the Hamiltonian (\ref{s002}) with the Bethe ansatz equation 
\[
\frac{\cosh \left( \lambda _j-\frac \eta 2\right) }{\cosh \left( \lambda _j+
\frac \eta 2\right) }\frac{\sinh ^2\left( \lambda _j+c+\frac \eta 2\right) }{
\sinh ^2\left( \lambda _j+c-\frac \eta 2\right) }\;\;\;\;\;\;\;\;\;\;\;\;\;
\; 
\]
\[
\cdot \frac{\sinh ^2\left( \lambda _j-c+\frac \eta 2\right) }{\sinh ^2\left(
\lambda _j-c-\frac \eta 2\right) }\left\{ \frac{\sinh \left( \lambda _j+
\frac \eta 2\right) }{\sinh \left( \lambda _j-\frac \eta 2\right) }\right\}
^{2N+1}\;\;\;\;\;\; 
\]
\begin{equation}
=-\prod_{k=1}^M\frac{\sinh \left( \lambda _j-\lambda _k+\eta \right) }{\sinh
\left( \lambda _j-\lambda _k-\eta \right) }\frac{\sinh \left( \lambda
_j+\lambda _k+\eta \right) }{\sinh \left( \lambda _j+\lambda _k-\eta \right) 
}.  \label{s001}
\end{equation}
where $j=1,2,\cdots ,M$ with $M$ being the number of down spins. Let\cite
{papa01,papa02} 
\begin{eqnarray*}
\Phi \left( \lambda ,\frac \gamma 2\right) &=&2\tan ^{-1}\left[ \cot \frac 
\gamma 2\tanh \lambda \right] ,\quad \\
e^{2i\Gamma } &=&\frac{1-e^{2i\gamma }}{e^{i\gamma }-e^{-i\gamma }}.
\end{eqnarray*}
By taking the logarithm of the Bethe ansatz equation (\ref{s001}), we have
that 
\begin{eqnarray}
&&2\Phi \left( \lambda _j+c,\frac \gamma 2\right) +2\Phi \left( \lambda _j-c,
\frac \gamma 2\right)  \nonumber \\
&&+\left( 2N+1\right) \Phi \left( \lambda _j,\frac \gamma 2\right) -\Phi
\left( \lambda _j,\Gamma \right)  \nonumber \\
&=&4\pi I_j+\sum_{k=1}^M\left\{ \Phi \left( \lambda _j-\lambda _k,\gamma
\right) +\Phi \left( \lambda _j+\lambda _k,\gamma \right) \right\}
\label{s006}
\end{eqnarray}
where $\eta =i\gamma $ . $I_j$ is an integer or the half-odd integer. Under
the thermodynamic limits we can write down the Bethe ansatz equation as the
form 
\begin{eqnarray}
&&\int_{-\infty }^\infty d\mu \sigma \left( \mu \right) \Phi ^{\prime
}\left( \lambda -\mu ,\gamma \right) +2\pi \sigma \left( \lambda \right) 
\nonumber \\
&=&\Phi ^{\prime }\left( \lambda ,\frac \gamma 2\right) +\frac 1{2N}\sigma
^N\left( \lambda \right)  \label{s005}
\end{eqnarray}
where 
\begin{eqnarray*}
\sigma ^N\left( \lambda \right) &=&2\Phi ^{\prime }\left( \lambda +c,\frac 
\gamma 2\right) +2\Phi ^{\prime }\left( \lambda -c,\frac \gamma 2\right) \\
&&-\Phi ^{\prime }\left( \lambda ,\Gamma \right) +\Phi ^{\prime }\left(
\lambda ,\frac \gamma 2\right) .
\end{eqnarray*}
The $\sigma \left( \lambda \right) $ is the distributed function. The $
\sigma ^N\left( \lambda \right) $ is the effect of the open boundary
condition and the magnetic impurities. It will give us the finite size
corrections introduced by the two impurities coupled to both of the ends of
the system. Fourier transformation on the equation (\ref{s005}) gives us the
distributed function as the form 
\begin{eqnarray*}
\sigma \left( \lambda \right) &=&\frac 1{2\gamma \cosh \frac{\pi \lambda }
\gamma }+\frac 1{2N}\frac 1{\left( 2\pi \right) ^2}\cdot \\
&&\cdot \int_{-\infty }^\infty d\omega e^{-i2\lambda \omega }\frac{\sinh
\left( \pi \omega \right) \widetilde{\sigma }^N\left( \omega \right) }{
2\sinh \left( \pi \omega -\gamma \omega \right) \cosh \left( \gamma \omega
\right) }
\end{eqnarray*}
with 
\begin{eqnarray*}
\widetilde{\sigma }^N\left( \omega \right) &=&\frac{4\pi \sinh \left( \pi
\omega -\gamma \omega \right) }{\sinh \left( \pi \omega \right) }\cdot \\
&&\cdot \left\{ 1+4\cos \left( 2c\omega \right) -\frac{\sinh \left( \pi
\omega -2\Gamma \omega \right) }{\sinh \left( \pi \omega -\gamma \omega
\right) }\right\}
\end{eqnarray*}
where the second term with the factor $1/\left( 2N\right) $ denotes the
finite size correction due to open boundary impurities. Therefore, the
distributed function has the form 
\[
\sigma \left( \lambda \right) =\frac 1{2\gamma \cosh \frac{\pi \lambda }
\gamma }+\frac 1{2N}\sigma _f\left( \lambda \right) 
\]
with the finite correction: 
\begin{eqnarray*}
\sigma _f\left( \lambda \right) &=&\frac{-1}{2\pi }\int_{-\infty }^\infty
d\omega e^{-i2\lambda \omega }\frac{\sinh \left( \pi \omega -2\Gamma \omega
\right) }{\sinh \left( \pi \omega -\gamma \omega \right) \cosh \left( \gamma
\omega \right) } \\
&&+\frac 1{\gamma \cosh \frac{\pi \left( \lambda +c\right) }\gamma }+\frac 1{
\gamma \cosh \frac{\pi \left( \lambda -c\right) }\gamma } \\
&&+\frac 1{2\gamma \cosh \frac{\pi \lambda }\gamma }.
\end{eqnarray*}
It is owing to the magnetic impurities coupled to the ends of the spin
chain. The ground state energy per site for our system is 
\begin{eqnarray*}
\frac EN &=&2\sin ^2\gamma \int_{-\infty }^\infty \frac{d\lambda }{\cosh
\left( \pi \lambda \right) \left[ \cos \gamma -\cosh \left( 2\gamma \lambda
\right) \right] } \\
&&+\cos \gamma +\frac 1N\left( \frac{2J}{\cosh c}-1\right. \\
&&\left. +2\int_{-\infty }^\infty d\lambda \frac{\sigma _f\left( \lambda
\right) \sin ^2\gamma }{\cos \gamma -\cosh \left( 2\lambda \right) }\right) .
\end{eqnarray*}
It is obvious that the ground state energy corresponding to the bulk is
exactly same as the periodic case\cite{yyang}. Its finite correction is,
based on the boundary condition, 
\begin{eqnarray*}
E_b &=&\left| \sin \gamma \right| \int_{-\infty }^\infty d\omega \frac{\sinh
\left[ \left( \pi -2\Gamma \right) \omega \right] }{\sinh \left( \pi \omega
\right) \cosh \left( \gamma \omega \right) }-1 \\
&&+\int_{-\infty }^\infty d\lambda \frac{\sin ^2\gamma }{\cosh \left( \pi
\lambda \right) \left[ \cos \gamma -\cosh \left( 2\gamma \lambda \right)
\right] }.
\end{eqnarray*}
The contribution given by the magnetic impurities has the following form: 
\begin{eqnarray*}
E_i &=&\int_{-\infty }^\infty d\lambda \frac{4\sin ^2\gamma }{\cosh \left(
\pi \lambda \right) \left[ \cos \gamma -\cosh \left( 2\gamma \lambda
-2c\right) \right] } \\
&&+\frac{2\sin ^2\gamma }{\sinh ^2c+\sin ^2\gamma }.
\end{eqnarray*}
Obviously, when $c=0,$ our model degenerates to the $XXZ$ spin chain with
the free boundary condition.

In the following section, we shall study 
further the physical properties of the
impurity model using the Luttinger-Fermi theory. We can regard $2\pi
I_j/N$ in equation (\ref{s006}) as the momentum of the $j$th quasi-particle.
Similarly as in the periodic boundary case, 
the energy $E$ takes its minimized
value when $\lambda \rightarrow \infty $ . Then the Fermi velocity of the
system can be defined as 
\[
v_F=\left. \frac{\varepsilon _d^{\prime }\left( \lambda \right) }{2\pi
\sigma \left( \lambda \right) }\right| _{\lambda \rightarrow \infty } 
\]
where the symbol $\varepsilon _d$ denotes the dressed energy of the model
(See reference [29]). Now we set $\sigma _f=\sigma _i+\sigma _b$ with
the relations 
\[
\sigma _i\left( \lambda \right) =\frac 1{\gamma \cosh \frac{\pi \left(
\lambda +c\right) }\gamma }+\frac 1{\gamma \cosh \frac{\pi \left( \lambda
-c\right) }\gamma }, 
\]
\begin{eqnarray*}
\sigma _b\left( \lambda \right) &=&\frac{-1}{2\pi }\int_{-\infty }^\infty
d\omega \frac{e^{-i2\lambda \omega }\sinh \left( \pi \omega -2\Gamma \omega
\right) }{\sinh \left( \pi \omega -\gamma \omega \right) \cosh \left( \gamma
\omega \right) } \\
&&+\frac 1{2\gamma \cosh \frac{\pi \lambda }\gamma },
\end{eqnarray*}
describing the finite corrections of the distributed function contributed by
the magnetic impurities and open boundary condition, respectively. The state
density of the system at the Fermi surface can be expressed as\cite
{mmm01,mmm02,mmm03,many05} 
\begin{eqnarray*}
N\left( \infty \right) &=&\frac 1{2\pi v_F}\left\{ 1+\frac 1{2N}\frac{\sigma
_i\left( \lambda \right) }{\sigma _0\left( \lambda \right) }\right. \\
&&\left. \left. +\frac 1{2N}\frac{\sigma _b\left( \lambda \right) }{\sigma
_0\left( \lambda \right) }\right\} \right| _{\lambda \rightarrow \infty }.
\end{eqnarray*}
Therefore, the low temperature specific heat contributed by the
magnetic impurities has the expression: 
\[
C_i=\frac 1{3N}\frac{\gamma \cosh \frac{\pi c}\gamma }{\sin \gamma }T. 
\]
where $T$ denotes the temperature of the system. The correction to the
susceptibility is 
\[
\chi _i=\frac 4N\cosh \left( \frac{\pi c}\gamma \right) \chi _0 
\]
where $\chi _0$ is the susceptibility in the bulk. The Kondo temperature
for our system is given by 
\[
T_k=\frac N{2\pi }\frac 1{\cosh \frac{\pi c}\gamma } 
\]
since the Kondo temperature is nothing but the effective Fermi
temperature in the local Fermi-liquid theory\cite{wanglett}. Notice that the
Kondo temperature $T_k\sim N\cos ^{-1}\left( \pi c/\gamma \right) $ when $
c\rightarrow ic$. This is similar to the result for the $XXX$ Heisenberg
spin chain and supports also the result obtained earlier in Ref. [34], 
that there is a crossover from an exponential law to an algebraic
law when the coupling constant between the host spins and the impurities
changes for the Kondo temperature.

In the above discussion, the impurity contributions are obtained in the
thermodynamic limits since the Bethe ansatz equations can be written down as
the integral equations which may be solved by a Fourier transformation. It
is an interesting problem to discuss the impurity effects for the finite
lattice case. In fact, in Ref. [35] de Vega and Woynarovich have
given a method to calculate the leading-order finite size corrections to the
ground state energy of the model which is soluble by the Bethe ansatz. The
situations for the $XXZ$ Heisenberg chain and the Hubbard chain are also
studied\cite{ham01,ham02,2996} by using the finite size scaling technique 
\cite{11301,11302,11303}. The impurity effects in different sectors of the
present model and its critical properties are also the interesting subjects
for further discussions.

To conclude, we have studied the solvable magnetic impurity model within the 
framework of open boundary $XXZ$ Heisenberg chain where the impurities are
coupled to the ends of the system. The Hamiltonian of the system with
impurities can been derived from the transfer matrix. The finite size
correction to the ground state energy is obtained for the contribution of
the magnetic impurities. With the help of the Luttinger-Fermi description we
get the specific heat and the susceptibility due to the impurities. The 
Kondo temperature is also given explicitly.

%%\end{multicols}


\begin{references}
\bibitem{yp6}  N. Andrei, K. Furuya, and J.H. Lowenstein, Rev. Mod. Phys. 
{\bf 55}, 331 (1983).

\bibitem{che3}  C.L. Kane and M.P.A. Fisher, Phys. Rev. Lett. {\bf 68}, 1220
(1992); Phys. Rev. B {\bf 46,} 15233 (1992).

\bibitem{che4}  I. Affleck, Nucl. Phys. B {\bf 336} 517 (1990).

\bibitem{a001}  A. Furusaki and N. Nagaosa, Phys. Rev. Lett. {\bf 72,} 892
(1994).

\bibitem{a002}  P. Fr$\stackrel{\text{..}}{\text{o}}$jdh and H. Johannesson,
Phys. Rev. Lett. {\bf 75}, 300 (1995).

\bibitem{many01}  G. Bed$\stackrel{\text{..}}{\text{u}}$rftig, F.H.L. E$
\beta $ ler, H. Frahm, Phys. Rev. Lett. {\bf 69}, 5098 (1996); Nucl. Phys. B 
{\bf 489,} 697 (1997).

\bibitem{many02}  P. Schlottmann and A.A. Zvyagin, Phys. Rev. B {\bf 55,}
5027 (1997).

\bibitem{many03}  A.A. Zvyagin and P. Schlottmann, Phys. Rev. B {\bf 56,}
300 (1997).

\bibitem{many04}  A.A. Zvyagin, Phys. Rev. Lett. {\bf 79}, 4641 (1997).

\bibitem{2701}  A.M. Tsvelick and P.B. Wiegmann, Adv. Phys. {\bf 32}, 453
(1983).

\bibitem{2702}  P. Schlottmann, Phys. Rep. {\bf 181}, 1 (1989).

\bibitem{2703}  P. Schlottmann and P.D. Sacramento, Adv. Phys. {\bf 42}, 641
(1993).

\bibitem{777}  N. Andrei and H. Johnesson, Phys. Lett. {\bf 100A}, 108
(1984).

\bibitem{888}  K.J.B. Lee and P. Schlottmann, Phys. Rev. B {\bf 37}, 379
(1988); P. Schlottmann, J. Phys. Condens. Matter {\bf 3}, 6617 (1991).

\bibitem{many05}  Y. Wang and J. Voigt, Phys. Rev. Lett. {\bf 77,} 4934
(1996).

\bibitem{many06}  Y. Wang, J. Dai, Z.N. Hu and F.C. Pu, Phys. Rev. Lett. 
{\bf 79,} 1901 (1997).

\bibitem{many07}  Z.N. Hu, F.C. Pu, and Y. Wang, Integrabilities of the $t-J$
Model with Impurities, to appear in J. Phys. A.

\bibitem{many08}  Z.N. Hu and F.C. Pu, Effects of Magnetic Impurities in the 
$t-J$ Model, Preprint.

\bibitem{his01}  E.K. Sklyanin, J. Phys. A {\bf 21}, 2375 (1988).

\bibitem{his02}  L. Mezincescu, R.I. Nepomechie, and V. Rittenberg, Phys.
Lett. A {\bf 147}, 70 (1990).

\bibitem{his03}  A. Foerster and M. Karowski, Nucl. Phys. B {\bf 408}, 512
(1993).

\bibitem{his04}  H.J. de Vega and A. Gonz$\stackrel{^{\prime }}{a}$lez-Ruiz,
Nucl. Phys. B {\bf 417,} 553 (1994).

\bibitem{wangonly}  Y. Wang, Phys. Rev. B {\bf 56}, 14045 (1997).

\bibitem{chen01}  S. Chen, Y. Wang, and F.C. Pu (unpublished).

\bibitem{boook}  V.E. Korepin, A.G. Izergin, and N.M. Bogoliubov, {\it 
Quantum Inverse Scattering Method, Correlation Functions and Algebraic Bethe
Ansatz }(Cambridge University Press, Cambridge, 1993).

\bibitem{papa01}  C.J. Hamer, G.R.W. Quispel, and M.T. Batchelor, J. Phys. A 
{\bf 20}, 5677 (1987).

\bibitem{papa02}  F.C. Alcaraz, M.N. Barber, M.T. Batchelor, R.J. Baxter,
and G.R.W. Quispel, J. Phys. A {\bf 20}, 6397 (1987).

\bibitem{yyang}  C.N. Yang and C.P. Yang, Phys. Rev. {\bf 150}, 321 (1966).

\bibitem{2996}  H. Asakawa and M. Suzuki, J. Phys. A {\bf 29}, 225 (1996).

\bibitem{mmm01}  J.M. Luttinger, J. Math. Phys. {\bf 4}, 1154 (1963); D.C.
Mattis and E.H. Lieb, J. Math. Phys. {\bf 6}, 304 (1965); F.D. Haldane, {\it 
ibid}. {\bf 45}, 1358 (1980).

\bibitem{mmm02}  P. Nozi$\stackrel{\text{`}}{\text{e}}$res, J. Low Temp.
Phys. {\bf 17}, 31 (1974).

\bibitem{mmm03}  J. Carmelo and A.A. Ovchinnikov, J. Phys. Condens. Matter 
{\bf 3}, 757 (1991); J. Carmelo, $et$ $al$., Phys. Rev. B {\bf 44,} 9967
(1991).

\bibitem{wanglett}  Y. Wang, Fermi-liquid description for the
one-dimensional electron systems, to appear in Phys. Rev. Lett..

\bibitem{Leet}  D.-H. Lee and J. Toner, Phys. Rev. Lett. {\bf 69,} 3378
(1992).

\bibitem{19855}  H.J. de Vega and F. Woynarovich, Nucl. Phys. B {\bf 251},
439 (1985).

\bibitem{ham01}  C.J. Hamer, J. Phys. A {\bf 19}, 3335 (1986).

\bibitem{ham02}  F. Woynarovich, J. Phys. A {\bf 22}, 4243 (1989).

\bibitem{11301}  J.L. Cardy, Nucl. Phys. B {\bf 270}, 186 (1986).

\bibitem{11302}  H.W.J. Bl$\stackrel{^{\prime \prime }}{o}$te, J.L. Cardy,
and M.P. Nightingale, Phys. Rev. Lett. {\bf 56}, 742 (1986).

\bibitem{11303}  I. Affleck, Phys. Rev. Lett. {\bf 56}, 746 (1986).
\end{references}
\end{document}